\documentclass[]{iaus}

\usepackage{graphicx}
\usepackage{aas_macros}
\usepackage{natbib}

\title[The Centre of M83] 
{The Centre of M83} 

\author[R. C. W. Houghton \& N. Thatte]   
{Ryan C. W. Houghton$^1$%
,
N. Thatte$^1$}

\affiliation{$^1$Astrophysics, University of Oxford, Denys Wilkinson Building, Oxford, OX1 3RH, U.K.\\[\affilskip]
}

\pubyear{2007}
\volume{245}  
\pagerange{119--126}
\date{?? and in revised form ??}
\jname{IAU Symposium}
\editors{M. Bureau (chief editor), E. Athanassoula, \& B. Barbuy., eds}
\begin{document}

\maketitle

\begin{abstract}
Stellar kinematics show no evidence of hidden mass concentrations at the centre of M83. We show the clearest evidence yet of an age gradient along the starburst arc and interpret the arc to have formed from orbital motion away from  a starforming region in the dust lane.

\end{abstract}

\firstsection 
\section{Introduction}\label{sec:intro}
The nucleus of M83 is offset from the bulge centre and surrounded by a semicircular starburst arc (Fig. 1). \citet[][hereafter H01]{Harris01} dated the star clusters in the arc with WFPC2 photometry and found evidence of an age gradients along it. However, the reddening vector parallelled the tracks in the two-colour diagram and clusters may be overlooked or poorly sampled in the visible because of extinction. \citet[hereafter T00]{Thatte00} proposed the existence of a second mass concentration after NIR long-slit kinematics revealed an additional peak in the velocity dispersion, 2.7\hbox{$^{\prime\prime}$}\ SW of the nucleus. Further studies with IFUs linked velocity gradients in gas kinematics to hidden mass concentrations at different locations: \citet{Mast06} report a gradient in \textrm{H}{\large $\alpha$}\ at 3\hbox{$.\!\!^{\prime\prime}$}9$\pm$0\hbox{$.\!\!^{\prime\prime}$}5 W of the nucleus while Diaz et al. (\citeyear[hereafter D06a and D06b]{D06a,D06b}) report a gradient in \textrm{Pa}{$\beta$}\ 7\hbox{$^{\prime\prime}$}\ WNW of the nucleus. However, gas kinematics are known to suffer non-gravitational effects \citep{KR95}.

We have analysed new VLT ISAAC K band long-slit spectroscopy together with archival HST data (\textrm{Pa}{\large $\alpha$}\ from NICMOS and \textrm{H}{\large $\alpha$}\ from WFPC2). Fig. 1 illustrates slit positions on the HST data. The combined data gives equivalent width (EW) measurements of \textrm{H}{\large $\alpha$}, \textrm{Pa}{\large $\alpha$}\ and CO (2.3\hbox{$\umu$m}), as well as stellar and gas kinematics.

\begin{figure}
\centering
\begin{minipage}[t]{0.7\textwidth}
\centering
\resizebox{0.9\textwidth}{!}{\includegraphics{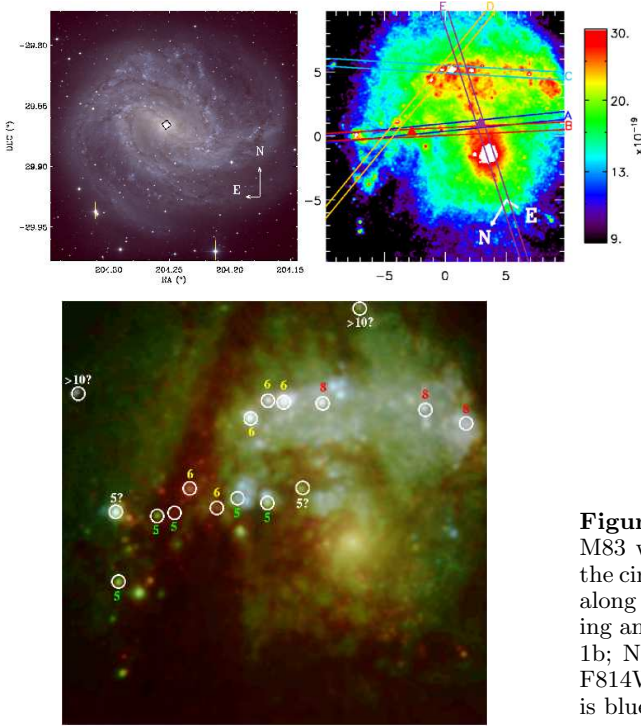} }
\end{minipage}%
\begin{minipage}[c]{0.3\textwidth}
\vspace{-1.5cm}
\caption[]{\textbf{(a)} A BVR image of M83 (logarithmic intensity) using data from \citet{Larsen99}, with the footprint of the homogenised HST data overlaid. Axes are scaled in degrees. \textbf{(b)} F222N NICMOS image of the central 20\hbox{$^{\prime\prime}$}$\times$20\hbox{$^{\prime\prime}$} with ISAAC slit positions overlaid. Positions of the putative hidden mass concentrations are also shown as purple (T00) and red (D06a,D06b) triangles. Flux is given in ergs s$^{-1}$ cm$^{-2}$ and axes are scaled in arcseconds. }
\end{minipage}

\label{fig:m83}
\end{figure}

\begin{figure}
\centering
\begin{minipage}[r]{0.6\textwidth}
	\centering
	\vspace{-2.5cm}
	\includegraphics[width=0.7\textwidth]{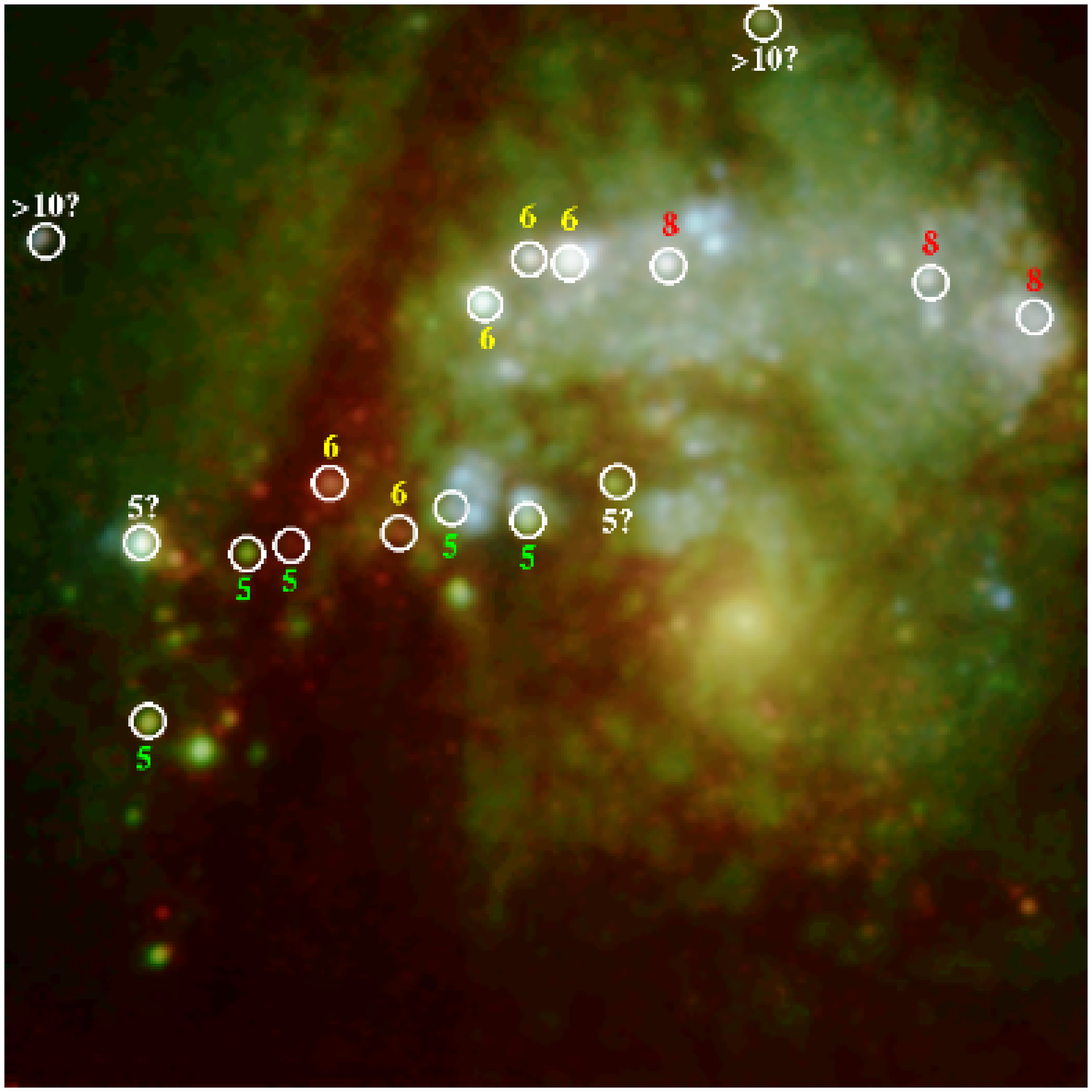}
\end{minipage}%
\begin{minipage}[c]{0.4\textwidth}
\caption[]{An image of the centre of M83 with ages overplotted; note that the circumnuclear arc extends into and along the dust lane; the position, scaling and orientation is the same as Fig. 1b; NICMOS F222M is red, WFPC2 F814W is green and WFPC2 F300W is blue.}
\end{minipage}
	
\label{fig:sb99}
\end{figure}
\section{Results}\label{sec:agedating}
To date the star clusters, we use the EWs of the CO bandhead and the \textrm{H}{\large $\alpha$}\ and \textrm{Pa}{\large $\alpha$}\ emission.
We applied Calzetti's extinction law \citep{Calzetti01} to correct the \textrm{H}{\large $\alpha$}\ and \textrm{Pa}{\large $\alpha$}\ EWs. Instantaneous SSP models \citep[hereafter SB99]{SB99} for \textrm{H}{\large $\alpha$}, \textrm{Pa}{\large $\alpha$}\ and CO were incompatible with the data for individual clusters and we investigated the subsequent evolution of a finite episode of star formation by convolving SSP model fluxes with a top hat kernel; we found an episode of 6 Myrs fits the data and gives the clearest evidence yet of an age gradient along the arc (Fig. 2).

We see a gradient in the H$_2$ (2-1 S(1)) gas velocity at the same vicinity as D06a report a gradient in \textrm{Pa}{\large $\alpha$}\ (Fig. 3). However, the stellar velocity dispersion shows no peak indicative of a mass concentration; we therefore attribute the gradients to a shock on the inner edge of dust lane, as predicted by \citet{Athanassoula92}. At the position reported by T00, we also see no clear evidence of a dispersion peak along Slit A.

\citet{Athanassoula92} showed that dust-lanes in bars are a consequence of gas on x1 orbits shocking and falling onto x2 orbits. Star formation in the dust lane would usually be opposed due to the shear forces in action along the shocks but we observe the shock front (located by the velocity gradient in the gas) on the inner edge of the dust lane. On the outer edge, away from the shock front, where there is likely to be spurs and feathering \citep{BD06}, star formation may be less opposed. Star clusters formed here would fall onto x2 orbits to produce the `star forming arc' and the age gradient.
\begin{figure}[h]
\centering
\begin{minipage}[c]{0.5\textwidth}
	\includegraphics[width=\textwidth,height=0.8cm]{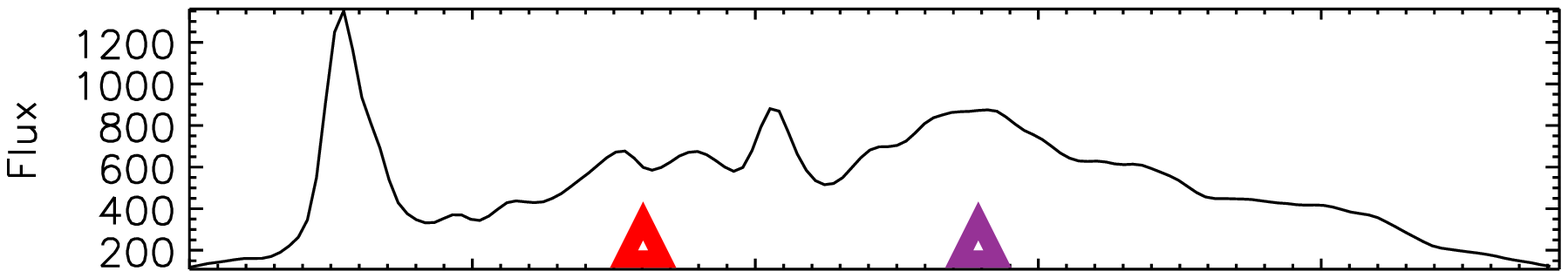}
	\includegraphics[width=\textwidth,height=2.2cm]{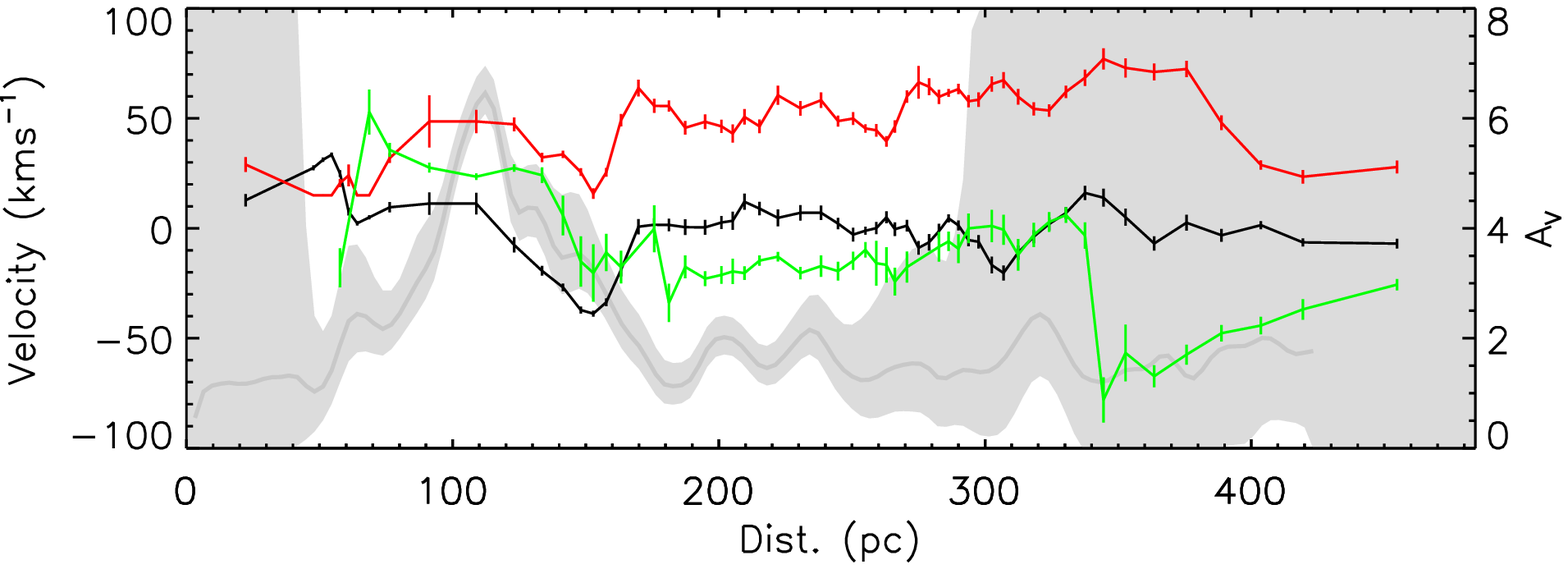}
\end{minipage}%
\begin{minipage}[c]{0.5\textwidth}
\caption[]{Kinematics along slit A.  Positions of the putative mass concentrations are shown as purple (T00) and red (D06a, D06b) triangles. The black, red and green lines illustrate the stellar velocity (mean subtracted), stellar velocity dispersion and the H$_2$ gas velocity, respectively; the grey line shows $\textrm{A}_{\textrm{\scriptsize V}}$\ and the lighter grey shading illustrates a $2\sigma$ error.}
\end{minipage}
\label{fig:slitAkin}
\end{figure}

\vspace{-0.9cm}
\bibliographystyle{mn2e}
\bibliography{HOUGHTON.bib}

\begin{thebibliography}{11}
\expandafter\ifx\csname natexlab\endcsname\relax\def\natexlab#1{#1}\fi

\bibitem[{{Athanassoula}(1992)}]{Athanassoula92}
{Athanassoula} E., 1992, \mnras, 259, 345

\bibitem[{{Bonnell} {et~al.}(2006){Bonnell}, {Dobbs}, {Robitaille}, \&
  {Pringle}}]{BD06}
{Bonnell} I.~A., {Dobbs} C.~L., {Robitaille} T.~P., {Pringle} J.~E., 2006,
  \mnras, 365, 37

\bibitem[{{Calzetti}(2001)}]{Calzetti01}
{Calzetti} D., 2001, \pasp, 113, 1449

\bibitem[{{D{\'{\i}}az} {et~al.}(2006b){D{\'{\i}}az}, {Dottori}, {Aguero},
  {Mediavilla}, {Rodrigues}, \& {Mast}}]{D06b}
{D{\'{\i}}az} R.~J., {Dottori} H., {Aguero} M.~P., {Mediavilla} E., {Rodrigues}
  I., {Mast} D., 2006b, \apj, 652, 1122

\bibitem[{{Diaz} {et~al.}(2006a){Diaz}, {Dottori}, {Mediavilla}, {Aguero}, \&
  {Mast}}]{D06a}
{Diaz} R.~J., {Dottori} H., {Mediavilla} E., {Aguero} M., {Mast} D., 2006a, New
  Astronomy Review, 49, 547

\bibitem[{{Harris} {et~al.}(2001){Harris}, {Calzetti}, {Gallagher},
  {Conselice}, \& {Smith}}]{Harris01}
{Harris} J., {Calzetti} D., {Gallagher} III J.~S., {Conselice} C.~J., {Smith}
  D.~A., 2001, \aj, 122, 3046

\bibitem[{{Kormendy} \& {Richstone}(1995)}]{KR95}
{Kormendy} J., {Richstone} D., 1995, \araa, 33, 581

\bibitem[{{Larsen} \& {Richtler}(1999)}]{Larsen99}
{Larsen} S.~S., {Richtler} T., 1999, \aap, 345, 59

\bibitem[{{Leitherer} {et~al.}(1999){Leitherer}, {Schaerer}, {Goldader},
  {Delgado}, {Robert}, {Kune}, {de Mello}, {Devost}, \& {Heckman}}]{SB99}
{Leitherer} C., {Schaerer} D., {Goldader} J.~D., {Delgado} R.~M.~G., {Robert}
  C., {Kune} D.~F., {de Mello} D.~F., {Devost} D., {Heckman} T.~M., 1999,
  \apjs, 123, 3

\bibitem[{{Mast} {et~al.}(2006){Mast}, {D{\'{\i}}az}, \& {Ag{\"u}ero}}]{Mast06}
{Mast} D., {D{\'{\i}}az} R.~J., {Ag{\"u}ero} M.~P., 2006, \aj, 131, 1394

\bibitem[{{Thatte} {et~al.}(2000){Thatte}, {Tecza}, \& {Genzel}}]{Thatte00}
{Thatte} N., {Tecza} M., {Genzel} R., 2000, \aap, 364, L47

\end{thebibliography}
\end{document}